%% file: coronal.tex
\documentclass[useAMS,usenatbib]{mn2e}
\usepackage{graphicx,rotate,url,mathptmx,lscape,times,color,amssymb,fixltx2e}

\usepackage{aas_macros}

\input{macros.tex}

\input{LocalMacros.tex}

\title[Coronal line emission in NGC~4696]{
Implications of Coronal Line Emission in NGC~4696
\thanks{Contains material \copyright\ British Crown copyright 2011/MoD}}

\author[M. Chatzikos et al.]
{
	\parbox[]{6.2in}
	{
		M.       Chatzikos$^1$\thanks{E-mail:mchatzikos@gmail.com},
		R.~J.~R. Williams$^2$,
		G.~J.    Ferland$^1$,
		R.~E.~A. Canning$^{3,4}$,\\
		A.~C.    Fabian$^5$,
		J.~S.    Sanders$^{6}$,
		P.~A.~M. van Hoof$^7$,
		R.~M.    Johnstone$^5$,\\
		M.       Lykins$^{1}$,
		and
		R.~L.    Porter$^{8}$
		\\
		\footnotesize
		$^1$Department of Physics \& Astronomy, University of Kentucky, Lexington, KY 40506, USA\\
		$^2$AWE plc, Aldermaston, Reading RG7 4PR\\
		$^3$Kavli Institute for Particle Astrophysics and Cosmology (KIPAC), Stanford University, 452 Lomita Mall, Stanford, CA 94305-4085, USA\\
		$^4$Department of Physics, Stanford University, 452 Lomita Mall, Stanford, CA 94305-4085, USA\\
		$^5$Institute of Astronomy, University of Cambridge, Madingley Road, Cambridge CB3 0HA\\
		$^6$Max-Planck-Institut f\"{u}r extraterrestrische Physik (MPE), Giessenbachstrasse, D-85748 Garching, Germany\\
		$^7$Royal Observatory of Belgium, Ringlaan 3, 1180 Brussels, Belgium\\
		$^8$Department of Physics and Astronomy and Center for Simulational Physics, University of Georgia, USA
	}
}

\date{%Accepted .
      Received }

\pagerange{\pageref{firstpage}--\pageref{lastpage}}
\pubyear{}

\begin{document}

\maketitle

\label{firstpage}

\begin{abstract}
\noindent
We announce a new facility in the spectral code \cloudy{} that
enables
tracking the evolution of a cooling parcel of gas with time.
For gas cooling from temperatures relevant to galaxy clusters, earlier
calculations estimated the \emlineforb{Fe}{xiv} $\lambda$5303
/ \emlineforb{Fe}{x} $\lambda$6375 luminosity ratio, a critical
diagnostic of a cooling plasma, to slightly less than unity.
By contrast, our calculations predict a ratio $\sim$3.
We revisit recent optical coronal line observations along
the X-ray cool arc around NGC~4696 by \citet{CanningCoronal11},
which detected \emlineforb{Fe}{x} $\lambda$6375, but not
\emlineforb{Fe}{xiv} $\lambda$5303.
We show that these observations are not consistent with
predictions of cooling flow models.
Differential extinction could in principle account for the
observations, but it requires extinction levels ($A_V > 3.625$)
incompatible with previous observations.
The non-detection of \emlineforb{Fe}{xiv} implies
a temperature ceiling of 2.1 million K.
Assuming cylindrical geometry and transonic turbulent pressure support, we
estimate the gas mass at $\sim$1 million $M_\odot$.
The coronal gas is cooling isochorically.
We propose that the coronal gas has not condensed out of the intracluster
medium, but instead is the conductive or mixing interface between the X-ray
plume and the optical filaments.
We present a number of emission lines that may be pursued to test this
hypothesis and constrain the amount of intermediate temperature gas in
the system.
\end{abstract}


\section{Introduction}
\label{intro}

\par
The X-ray-emitting hot intracluster medium (ICM) at the centres of cool
core clusters of galaxies can have cooling times shorter than a Hubble
time \citep[see, e.g.,][]{FabianARAA2012}.
Heating, most likely from active galactic nuclei (AGN) feedback, prevents
catastrophic cooling.
However, the brightest cluster galaxies in these systems are observed to have
substantial quantities of cool and cold gas and star formation relative
to their non-cool core analogues and it is widely believed that some
residual cooling must occur. 

\par
If gas is cooling radiatively then observations of intermediate
temperature gas could probe the cooling rates in these systems.
\citet{Sanders+2008-Cen} reported a temperature floor of
$\sim$4~million K in their dispersive X-ray spectra of the
Centaurus cluster.
The detection of \emline{O}{vii} at much lower levels than predicted
by cooling flow models \citep{SandersFabian2011-stacked-o7} indicates
that a dearth of 1--3~million K gas is common among cool clusters.
If cooling at cluster centres is not impeded by heating mechanisms,
it may proceed non-radiatively, e.g., through mixing.

\par
Recently, \citet[][C11a]{CanningCoronal11} reported a detection of
\emlineforb{Fe}{x} coronal line emission in NGC~4696, the brightest
cluster galaxy in the nearby ($z = 0.0104$, 44.3 Mpc) Centaurus cluster.
This line is sensitive to $\sim$million K gas.
If the emission arises from cooling gas, their detection implies a high
mass deposition rate of $\sim$20 M$_{\odot}$ yr$^{-1}$.
However, they do not detect the higher excitation line of
\emlineforb{Fe}{xiv}.

\par
This paper builds upon advances in the spectral simulation code \cloudy,
which permit simulations where the ionization is not in equilibrium with
the local gas kinetic temperature (Chatzikos et al., in prep.).
\citet{Lykins-Chianti} describe the improvements
in the atomic physics which underlies such simulations.

\par
The outline of the paper is as follows.
In Sec.~\ref{sec:sims} we present a brief overview of
our numerical implementation of non-equilibrium cooling.
In Sec.~\ref{sec:comp-graney} we present modern calculations
of the cooling efficiencies for the emission lines pursued by
C11a, and emphasize the improvements since \citet[][GS90]{GraneySarazin1990}.
In Sec.~\ref{sec:optical-depth} we establish that \emlineforb{Fe}{x}
is optically thin.
In Sec.~\ref{sec:xray-connection} we update the mass deposition
rate through \emlineforb{Fe}{x}, and compare it to the mass
deposition rate through the X-rays of \citet{Sanders+2008-Cen}.
In Sec.~\ref{sec:cooling-condens} we use our \cloudy{} simulations
to rule out the possibility that the coronal line gas is a cooling
condensation of the ambient ICM.
Then, in Sec.~\ref{sec:FeXIV-constraints} we use the non-detection of
\emlineforb{Fe}{xiv} to constrain the temperature structure of the
coronal gas, and in Sec.~\ref{sec:properties} we obtain its basic
properties based on the detected \emlineforb{Fe}{x}.
We discuss our results in Sec.~\ref{sec:discussion},
and summarize in Sec.~\ref{sec:summary}.

\section{Spectral Simulations}\label{sec:sims}

\par
This section outlines how we compute the spectrum of a cooling
non-equilibrium parcel of gas.
The calculations presented here were performed with
version r9230 on the ``dyna'' development branch of
the spectral synthesis code \cloudy{} \citep{C13}.

\par
In all cases we begin with a hot gas in collisional ionization equilibrium
(CIE) at a preset temperature. 
We then allow the gas to freely cool, solving for the gas kinetic temperature,
the non-equilibrium distribution of ionization (NEI), the gas cooling, and its
spectrum.

\par
The implementation of time-dependent physics within \cloudy{} is discussed
in detail in a separate paper (Chatzikos et al., in prep.).
Briefly, our approach builds upon the infrastructure developed by
\citet{Henney2005} to treat the structure of steady-state ionization fronts.
In its normal mode of usage, \cloudy{} uses a hierarchy of solvers to treat
the coupled non-linear systems of ionization and level balance, electron
density, chemical equilibrium, temperature equilibrium and (where appropriate)
pressure equilibrium.
By including appropriate source and sink terms in each of these equilibrium solvers,
we showed that they could also be used to implement an implicit form of advance from
an initial state.
In our previous work this initial state was upstream in the flow; in the
present case, it is that at the previous time-step.

\par
The time-dependent ionization balance equations may be written as
\begin{equation}
	{dn_i\over dt} = \sum_j R_{ji} n_j - R_{ij} n_i	\,	.
\end{equation}
In the standard solver, the terms on the left hand side of this
expression are assumed to be zero.  In our time-dependent approach, we
discretize over time-steps $m$ and time-steps $\Delta t$, and use an implicit
time-step advance for stability as the full balance system is very
stiff.  The discretized equations for the state $n_i^{m+1}$ at
time-step $m+1$ are then
\begin{equation}
{n_i^{m+1}-n_i^m\over \Delta t} = \sum_j R_{ji} n_j^{m+1} - R_{ij} n_i^{m+1},
\end{equation}
where the first term on the left hand side is treated as an additional
sink term in the balance matrix.

\par
This approach differs from that used by \citet{GnatSternberg07},
which was based on the database of rates included in 
a version of \cloudy{} that dated from 2006.
These authors integrated the ionization rates using a separate ODE
solver package, and called \cloudy{} directly only to update the cooling
rate of the plasma based on the current ionization fractions.
The approach which we have adopted allows us to be fully self-consistent
in all the components of the time-step advance, and allows the implicit
equations for the advanced state of the system to be solved in
discrete physical components, rather than requiring a linear system to
be solved for all states of all species in parallel.
However, the constraints of working around the existing code have meant
that it is at present only possible to use a first-order method for the
time advance, so shorter time-steps are required to maintain the accuracy
of the solution.

\par
The implementation permits the temporal integration of line emissivities.
These are shown to be related to line emission in the classical cooling
flow model.
In particular, cumulative density-weighted line
emissivities\footnote{Enabled with the command \textbf{set cumulative mass}.}
are by definition equal to the total line emission in a multi-temperature cooling
flow,
\begin{eqnarray}
	F_\mathrm{line}^\prime	& = &	\int{dt^\prime \, j_\mathrm{line}(T^\prime) / \rho}						\nonumber	\\
				& = &	\frac{\alpha}{2} \, \frac{k_B}{\mu \, m_p} \,
				         	\int_{T}^{T_\mathrm{max}}
				         		{\frac{\Lambda_\mathrm{line}(T^\prime)}{\Lambda(T^\prime)} \, dT^\prime}	\nonumber	\\
				& = &	\Gamma(T, T_\mathrm{max})	\,	,
	\label{eqn:mass-integral}
\end{eqnarray}
where $\alpha = 5$ or 3, for isobaric (constant pressure, CP) or
isochoric (constant density, CD) cooling, respectively,
and the mean molecular weight, $\mu$, is a slow function of temperature.
$\Lambda_\mathrm{line}(T)$ is the frequency-integrated line cooling
efficiency, while $\Lambda(T)$ is the cooling function.
The integral extends over the temperature range where $j_\mathrm{line}(T)$
in finite, but it is customary to extend the range from 0\unit{K} to the
highest temperature in the gas (and drop the zero from the notation for $\Gamma$).
Then, $\Gamma(T_\mathrm{max})$ encapsulates the total line emission per unit mass of
the multi-temperature cooling gas, and determines the line luminosity through
\begin{equation}
	L_\mathrm{line} = \dot{M} \, \Gamma(T_\mathrm{max})	\,	,
	\label{eqn:cflow-lum}
\end{equation}
where $\dot{M}$ is the mass deposition rate \citep[e.g.,][]{SarazinGraney91}.
The remaining symbols have their usual meaning.

\par
The previous equation was effected by employing the differential cooling time,
\begin{equation}
	dt = \frac{\alpha}{2} \, \frac{n_\mathrm{tot} \, k_B \, dT}{n_e \, n_H \, \Lambda(T)}	\,	.
        \label{eqn:cooling-time}
\end{equation}
Notice that the temporal integration weighs against phases near the peak
of the cooling function.

\par
In addition, unweighted cumulative line emissivities\footnote{Enabled with the
command \textbf{set cumulative flux}.} may be computed as
\begin{eqnarray}
	F_\mathrm{line} & = &	\int{dt \, j_\mathrm{line}(T)}				\nonumber	\\
			& = &	\frac{\alpha}{2} \, \int_0^{T_\mathrm{max}}
					{\frac{k_B \, \rho}{\mu \, m_p} \,
					\frac{\Lambda_\mathrm{line}(T)}{\Lambda(T)} \, dT}	\,	.
	\label{eqn:flux-integral}
\end{eqnarray}
For isochoric cooling the cumulative integral becomes
\begin{equation}
	F_\mathrm{line}	 =	\rho \, \Gamma(T_\mathrm{max})	\,	,
	\label{eqn:flux-integral:isochoric}
\end{equation}
while for isobaric cooling it becomes
\begin{equation}
	F_\mathrm{line}  =	\frac{5 \, P}{2} \, \int_0^{T_\mathrm{max}}
					{\frac{\Lambda_\mathrm{line}(T)}{\Lambda(T)} \,
						\frac{dT}{T}}	\,	.
	\label{eqn:flux-integral:isobaric}
\end{equation}
If the line emits over a restricted range, the temperature may be taken out
of the integral as constant, and eqn.~(\ref{eqn:flux-integral:isochoric}) is
recovered.
However, some lines, such as \emline{N}{v} $\lambda$1243 (see below), are prominent
over a wide range of temperatures and the inverse temperature dependence
in the integral must be retained.

\section{Comparison with Previous Calculations}\label{sec:comp-graney}

\par
The numerical advances and the atomic data updates \citep{Lykins-Chianti}
will cause significant changes from the predictions
in the seminal paper of GS90.
The recombination rates are now significantly larger due to the incorporation
of recombination channels which were not known twenty years ago.
The atomic line database used to compute the cooling is far larger than
was available previously, due to our use of both Opacity Project and Chianti
line data.
Even for lines in common with \citet{SarazinGraney91}, the cooling
can be quite different due to our use of high-quality close coupling
collision rates, rather than the simpler rates that were available at that time.

\par
Table~\ref{table:cen-flux-ratios} lists the optical coronal lines
of interest.
To facilitate a direct comparison to the GS90 predictions
for these lines, we have computed their emissivity for a
unit volume of gas, as it cools from $8 \times 10^7\unit{K}$,
appropriate to the ICM, to $10^5\unit{K}$, adopting the same
abundances \citep{MeyerSolar79}.

%%%%%%%%%%%%%%%%%%%%%%%%%%%%%%%%%%%%%%%%%%%%%%%%%%%%%%%%%%%%%%%%%%%%%%%%%%%%%%%%%
%				TABLE						%
%%%%%%%%%%%%%%%%%%%%%%%%%%%%%%%%%%%%%%%%%%%%%%%%%%%%%%%%%%%%%%%%%%%%%%%%%%%%%%%%%
\begin{table}
	\centering
	\caption
	{
		C11a optical coronal lines in Centaurus.
		The first two columns give the spectroscopic label and wavelength,
		while the third presents the upper limits relative to the \emlineforb{Fe}{x}
		$\lambda$6375 flux.
	}
	\label{table:cen-flux-ratios}
	\begin{tabular}{|c|c|c|}
		\hline
		\hline
		Spectral Label	& $\lambda$ / \AA	& Flux Ratio	\\
		\hline
		\emlineforb{Fe}{xiv}	& 5303			& $<$1.29	\\
		\emlineforb{Ca}{xv}	& 5445			& $<$0.29	\\
		\emlineforb{Ca}{xv}	& 5695			& $<$1.09	\\
		\emlineforb{Ni}{xv}	& 6700			& $<$0.31	\\
		\hline
	\end{tabular}
\end{table}
%%%%%%%%%%%%%%%%%%%%%%%%%%%%%%%%%%%%%%%%%%%%%%%%%%%%%%%%%%%%%%%%%%%%%%%%%%%%%%%%%

\par
\begin{figure}
	\begin{centering}
	\includegraphics{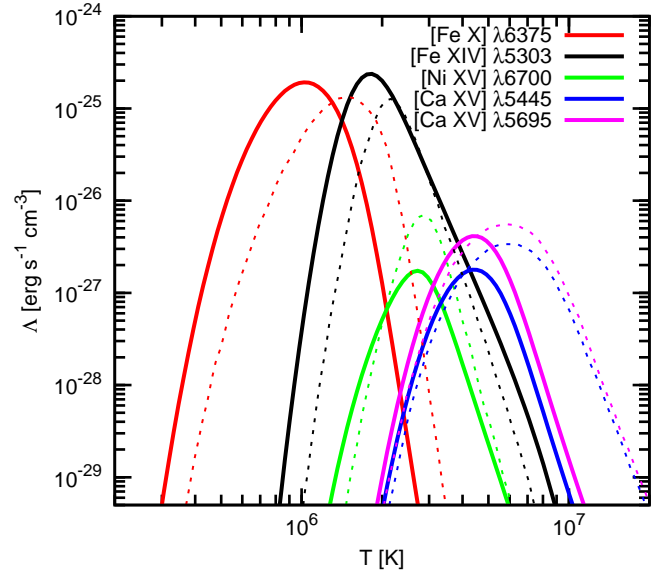}
	\end{centering}
	\caption[]
	{
		Isochoric cooling efficiency for the lines
		discussed by C11a.
		Solid and dotted lines show the present and GS90
		results, respectively.
		The lines as listed in the legend proceed from
		left to right.
	}
	\label{fig:graney90-comp}
\end{figure}

Figure~\ref{fig:graney90-comp} illustrates the comparison between
the emission coefficients reported by GS90
and the present calculations.
The most evident difference is a general shift to lower temperatures.
The calcium lines are suppressed by $\sim$1--2 dex above
8$\times$10$^6\unit{K}$.
The \emlineforb{Fe}{x} $\lambda$6375 line is shifted to a maximum around
a million K, leading to 1 dex of suppression at higher temperatures,
and a comparable boost at lower temperatures.
The cooling efficiency of \emlineforb{Fe}{xiv} $\lambda$5303 is also boosted
below its former peak, but is roughly unchanged above it.
The maximum efficiencies are within a factor of 2 of the GS90 results,
except \emlineforb{Ni}{xv} $\lambda$6700, which is depressed by $\sim$4.

\par
These remarks are relevant to the C11a observations.
The fact that the cooling efficiency of \emlineforb{Fe}{xiv} $\lambda$5303
has increased {\em over the entire temperature range} suggests that its
non-detection cannot be due to uncertainties in the atomic data, as they
cautioned.
Quite the contrary, the enhancement of the emissivity below $2 \times 10^6\unit{K}$
places more stringent limits on the exact temperature range covered by gas,
as we discuss below.

\section{\emlineforb{F\MakeLowercase{e}}{x} Optical Depth}\label{sec:optical-depth}

\par
\begin{figure}
	\begin{centering}
	\includegraphics{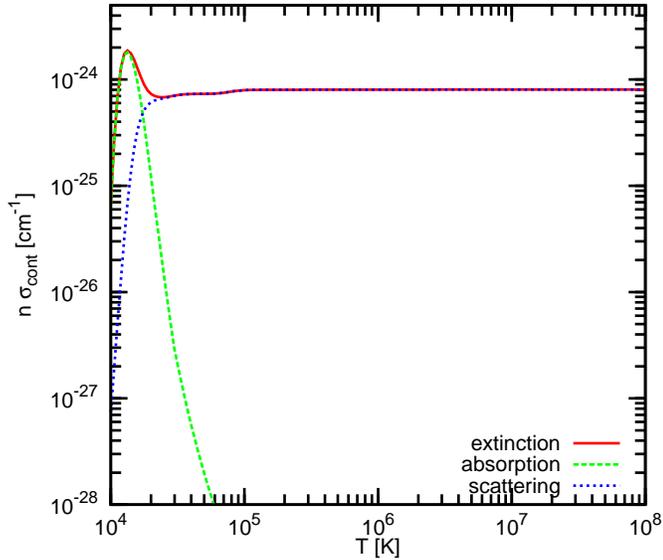}
	\end{centering}
	\caption[Continuum Opacity Temperature Dependence.]
	{
		Variation of continuum opacity with temperature
		for a 1\unit{cm} slab of unit hydrogen density.
		The contributions of absorption and scattering are
		also presented.
	}
	\label{fig:opacity}
\end{figure}

It is important to estimate the optical depth to \emlineforb{Fe}{x} through
the Centaurus galaxy cluster ICM, and the warm gas of NGC~4696.
Figure~\ref{fig:opacity} shows the optical depth due to
continuum opacity though a 1\unit{cm} slab of gas at unit hydrogen
density as a function of temperature.
The opacity is dominated by Thomson scattering off of free electrons.
Below 20,000\unit{K} the main contributor to opacity is continuum
absorption.

\par
Because of our choice of unit density, numerically the continuum opacity
is roughly equivalent to the extinction cross-section (to within a factor
of a few).
It follows that the inverse of the opacity amounts to the hydrogen column
density required for $\tau \sim 1$.
The largest value reported by \citet{Crawford2005-cen} is not higher
than 10$^{22}$\unit{cm^{-2}}.
This indicates that gas hotter than 10,000\unit{K} is optically
thin to radiation at \emlineforb{Fe}{x}.

\par
\citet{Mittal+2011-cen-herschel} computed the gas-to-dust ratio
through a 7\unit{kpc} aperture to be at best 70.
\citet{ODea+1994} estimated the amount of molecular gas to
$\sim$5$\times$10$^8$\unit{M_\odot} (see Mittal et al.), which we adopt here as
the total mass for the molecular and dust components.
This translates to a column density of $\sim$2$\times$10$^{21}$\unit{cm^{-2}}.
Combined with the declining absorption cross-section,
we conclude that the cold matter in NGC~4696 is also
optically thin to \emlineforb{Fe}{x}.

\par
Finally, extinction due to spectral lines is also negligible,
as it is at most five orders of magnitude smaller than the
continuum.

%%%%%%%%%%%%%%%%%%%%%%%%%%%%%%%%%%%%%%%%%%%%%%%%%%%%%%%%%%%%%%%%%%%%%%%%%%%%%%%%%
%				FIGURE						%
%%%%%%%%%%%%%%%%%%%%%%%%%%%%%%%%%%%%%%%%%%%%%%%%%%%%%%%%%%%%%%%%%%%%%%%%%%%%%%%%%
%%%%%%%%%%%%%%%%%%%%%%%%%%%%%%%%%%%%%%%%%%%%%%%%%%%%%%%%%%%%%%%%%%%%%%%%%%%%%%%%%

\section{Comparison to X-ray Gas}\label{sec:xray-connection}

\par
\citet{Sanders+2008-Cen} modelled their dispersive \xmm{} spectra
with cooling flow models and reported that no significant quantities
of cool ($\lesssim$0.4\unit{keV}) gas were detected.
Comparison with spatially resolved spectra obtained with \chandra{}
\citep{SandersFabian06} indicates that the lowest temperatures arise
from the central 30\arcsec{}, while higher temperatures are emitted
progressively farther from the cluster centre.
In fact, their best-fitting cooling flow models were consistent with mass
deposition rates that diminish with decreasing temperature.
An upper limit of 0.8\unit{M_\odot \, yr^{-1}} was obtained for
their lowest temperature of 0.25\unit{keV}.
Similar results were obtained from spectral line analysis.

\par
The coronal lines probe temperatures of 1--5~million K,
and offer a complementary view into cooling at and beyond
the softest bands accessible to X-rays.
In order to constrain the state and origin of the coronal
line gas, it is important to test whether it is consistent
with being due to cooling from a possibly abated cooling flow.

\par
For our analysis, we employ \cloudy{} calculations that track
a unit volume of gas at a hydrogen density of 0.07\unit{cm^{-3}}
as it cools from 80~million K to 10,000\unit{K}, and ignore the
transfer of the optically thin coronal lines.
(Our results are not sensitive to the exact density value.)
We adopt the abundances for the coolest of the five-component
cooling flow model of \citet{Sanders+2008-Cen} as the most
appropriate for the question at hand, listed in the third
column of Table~\ref{table:CentaurusAbundances-2008}.

%%%%%%%%%%%%%%%%%%%%%%%%%%%%%%%%%%%%%%%%%%%%%%%%%%%%%%%%%%%%%%%%%%%%%%%%%%%%%%%%%
%				TABLE						%
%%%%%%%%%%%%%%%%%%%%%%%%%%%%%%%%%%%%%%%%%%%%%%%%%%%%%%%%%%%%%%%%%%%%%%%%%%%%%%%%%
\begin{table}
	\centering
	\caption
	{
		Adopted elemental abundances relative to hydrogen
		in our simulations.
		The second column lists the solar abundances of
		\citet{AndersGrevesse1989}.
		The last two columns list the abundances for the cool
		gas component obtained in the fits of \citet{Sanders+2008-Cen}:
		the five-component cooling flow model (third column),
		and the five-temperature thermal model (last column).
	}
	\label{table:CentaurusAbundances-2008}
	\begin{tabular}{|l|c|c|c|c|}
		\hline
		\hline
		Element		&	Solar		&	5$\times$VMCFLOW&	5$\times$VAPEC	\\
		\hline
		Nitrogen	&	1.12E$-$4	&	3.93E$-$4	&	1.80E$-$4	\\
		Oxygen		&	8.51E$-$4	&	4.43E$-$4	&	4.00E$-$4	\\
		Neon		&	1.23E$-$5	&	9.84E$-$5	&	9.10E$-$5	\\
		Magnesium	&	3.80E$-$5	&	3.88E$-$5	&	3.46E$-$5	\\
		Silicon		&	3.55E$-$5	&	6.24E$-$5	&	5.68E$-$5	\\
		Calcium		&	2.29E$-$6	&	6.41E$-$6	&	5.96E$-$6	\\
		Iron		&	4.68E$-$5	&	5.05E$-$5	&	4.54E$-$5	\\
		Nickel		&	1.78E$-$6	&	4.52E$-$6	&	4.16E$-$6	\\
		\hline
	\end{tabular}
\end{table}
%%%%%%%%%%%%%%%%%%%%%%%%%%%%%%%%%%%%%%%%%%%%%%%%%%%%%%%%%%%%%%%%%%%%%%%%%%%%%%%%%

\par
\begin{figure}
	\begin{centering}
	\includegraphics{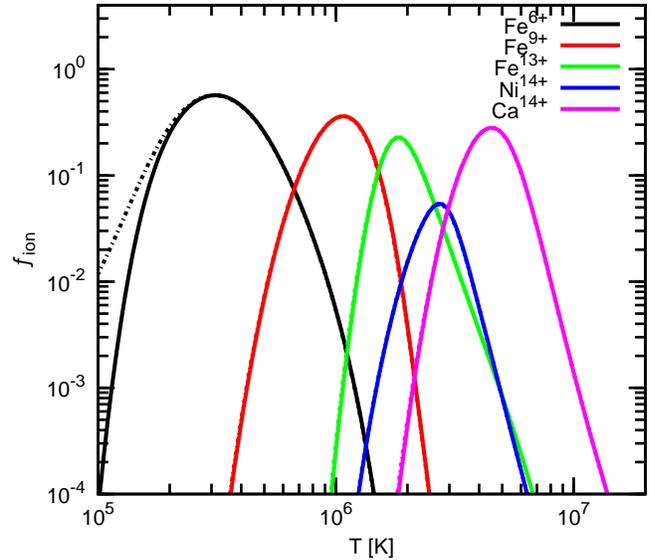}
	\end{centering}
	\caption[Ionization fractions as a function of temperature.]
	{
		Fractional densities for the ions discussed by C11a,
		and for Fe$^{6+}$.
		Results are presented for both CIE (solid) and NEI (dashed)
		calculations of isochoric cooling.
		Recombination lags become significant below $2 \times 10^5$\unit{K}.
		The ions as listed in the legend proceed from left to right.
	}
	\label{fig:ion-frac-vs-temp}
\end{figure}

Figure~\ref{fig:ion-frac-vs-temp} illustrates the ionization fractions
of the relevant ions for both CIE and NEI isochoric calculations.
Isobaric calculations produce identical results and are not shown.
Evidently, in this temperature range cooling and recombination are
in equilibrium, in agreement with GS90
and \citet{GnatSternberg07}.
The ionization fraction of Fe$^{6+}$ is also presented to illustrate
that non-equilibrium effects become important at lower temperatures.

%%%%%%%%%%%%%%%%%%%%%%%%%%%%%%%%%%%%%%%%%%%%%%%%%%%%%%%%%%%%%%%%%%%%%%%%%%%%%%%%%
%				FIGURE						%
%%%%%%%%%%%%%%%%%%%%%%%%%%%%%%%%%%%%%%%%%%%%%%%%%%%%%%%%%%%%%%%%%%%%%%%%%%%%%%%%%
%%%%%%%%%%%%%%%%%%%%%%%%%%%%%%%%%%%%%%%%%%%%%%%%%%%%%%%%%%%%%%%%%%%%%%%%%%%%%%%%%

\par
We begin by recalculating the mass deposition rate through
\emlineforb{Fe}{x}.
C11a reported that the luminosity in \emlineforb{Fe}{x}
is 3.5$\times$10$^{37}$\unit{erg \, s^{-1}}, which translates
to $\dot{M} \sim 1.63 \thinspace \Mdotunit$, given that
$\Gamma \approx 3.42 \times 10^{11}$\unit{erg \, g^{-1}}
for isobaric cooling.
(For isochoric cooling,
$\Gamma \approx 2.07 \times 10^{11}$\unit{erg \, g^{-1}},
and the mass rate is $\sim$2.68$\thinspace \Mdotunit$.)
This is significantly lower than the rate of 20$\thinspace \Mdotunit$,
deduced by C11a using the older results of \citet{SarazinGraney91},
and in overall agreement with the X-ray spectroscopy results.
Note, however, that this estimate is in excess of the upper limit
of 0.8$\thinspace \Mdotunit$ obtained by \citet{Sanders+2008-Cen}
by fitting the entire spectrum with a multi-component cooling flow
model.

\par
In Table~\ref{table:cen-xray-rates}, we investigate the consistency
of our cooling calculations against the cooling flow model predictions
of \citet[see Table 3 therein]{Sanders+2008-Cen}.
Following that work, our models cool from an initial temperature
of $\sim$4.5\unit{keV}.
However, we do not employ solar metallicities as in that work,
but instead use the abundances appropriate to their five-component
VMCFLOW model as before.
Note from Table~\ref{table:CentaurusAbundances-2008}
that the iron abundance is close to the solar value, while the nitrogen
abundance is 3.5 times higher.
We compute the $\Gamma(T)$ functions for these lines, and convert
the X-ray luminosities to mass deposition rates with the aid of
eqn.~(\ref{eqn:cflow-lum}).
Our calculations are in good agreement with the reported rates
(shown in the last column of Table~\ref{table:cen-xray-rates}).
\emline{N}{vii} is the only exception because
the adopted solar abundance in that paper is
unrealistically low for the X-ray gas (see Sanders
et al.{} for more details).
Adjusting the reported mass deposition rate through \emline{N}{vii}
by 3.5 leads to a value comparable to the rate calculated by \cloudy{}.

\par
Table~\ref{table:cen-xray-ratios} presents estimates for the mass
deposition rate through the \emlineforb{Fe}{x} coronal line.
These are obtained first by converting the ratio of X-ray to coronal
luminosities into ratios of the mass deposition rates, and then by
invoking the reported mass rates through the X-ray lines.
The mass deposition rates cover the range 1--4$\thinspace \Mdotunit$,
in agreement with our estimate from the coronal line alone.

\par
This analysis indicates that the mass deposition rate through the
coronal line is in reasonable agreement with the rates obtained from
the X-ray lines.
In principle, then, the coronal gas is consistent with cooling
out of the hot ICM at a uniform rate in the range 1--10 million K.
If that were the case, the expected luminosity for \emlineforb{Fe}{xiv}
is about 10$^{38}$\unit{erg \, s^{-1}}, or about twice the upper limit.
(This result may be obtained by comparing to \emline{Fe}{xvii} $\lambda$15.01
assuming a uniform deposition rate, and noting that for the coronal line
$\Gamma(T) = 5.9343$$\times$$10^{11}$\unit{erg \, g^{-1}} for isochoric cooling.)
The fact that the 2 million K \emlineforb{Fe}{xiv} line is not observed
is perplexing.

%%%%%%%%%%%%%%%%%%%%%%%%%%%%%%%%%%%%%%%%%%%%%%%%%%%%%%%%%%%%%%%%%%%%%%%%%%%%%%%%%
%				TABLE						%
%%%%%%%%%%%%%%%%%%%%%%%%%%%%%%%%%%%%%%%%%%%%%%%%%%%%%%%%%%%%%%%%%%%%%%%%%%%%%%%%%
\begin{table*}
	\centering
	\caption
	{
		Updated mass deposition rates for the detected X-ray lines of
		\citet{Sanders+2008-Cen} based on our isobaric calculations.
		For a more fair comparison, the integrations are done starting
		at a temperature of about 4.5\unit{keV}.
		However, we do not assume solar metallicities.
		Instead, our results are drawn from calculations employing the
		abundances of their 5$\times$VMCFLOW model (see
		Table~\ref{table:CentaurusAbundances-2008}).
		The first two columns identify the X-ray emission line.
		The third column lists the reported line luminosity.
		The fourth column lists the $\Gamma(T)$ functions,
		which are converted to mass deposition rates in
		the fifth column with the aid of eqn.~(\ref{eqn:cflow-lum}).
		Finally, the last column shows the reported mass
		deposition rates for comparison.
	}
	\label{table:cen-xray-rates}
	\begin{tabular}{|c|c|c|c|c|c|c|c|}
		\hline
		\hline
		Label					&
		$\lambda$ / \AA				&
		$L$ / 1E+39 $\mathrm{erg \, s^{-1}}$	&
		$\Gamma(T)$ / $\mathrm{erg \, g^{-1}}$	& 
		$\dot{M}_\mathrm{calc}$/ \Mdotunit{} 	&
		$\dot{M}_\mathrm{rep}$ / \Mdotunit{}	\\
		\hline
		\emline{Fe}{xvii}	&	15.01	& 12.4	& 1.23E+14	& 1.60	& 1.6	\\
		\emline{Fe}{xvii}	&	15.26	&  3.6	& 3.32E+13	& 1.72	& 1.6	\\
		\emline{Fe}{xvii}	&	16.78	&  4.3	& 6.13E+13	& 1.11	& 1.7	\\
		\emline{Fe}{xvii}	&	17.05	& 19.3	& 7.64E+13	& 4.00	& 3.0	\\
		\emline{N}{vii}		&	24.78	& 11.5	& 5.00E+13	& 3.65	& 9.7	\\
		\hline
	\end{tabular}
\end{table*}
%%%%%%%%%%%%%%%%%%%%%%%%%%%%%%%%%%%%%%%%%%%%%%%%%%%%%%%%%%%%%%%%%%%%%%%%%%%%%%%%%
\begin{table*}
	\centering
	\caption
	{
		Coronal mass deposition rates based on the detected X-ray lines
		reported by \citet{Sanders+2008-Cen}.
		The first two lines identify the X-ray line.
		The third column lists the ratio of the X-ray luminosity to that of
		\emlineforb{Fe}{x} $\lambda$6375.
		The fourth column lists the ratio of the $\Gamma(T)$ functions for a
		freely cooling gas (isochoric and isobaric cooling produce virtually
		identical ratios).
		The fifth column lists the ratio of mass deposition rates of the
		X-ray line relative to that of the coronal line, obtained with
		eqn.~(\ref{eqn:cflow-lum}).
		The last column lists the mass deposition rate in the coronal line
		using the reported X-ray mass deposition rate, listed in the last
		column of Table~\ref{table:cen-xray-rates}.
	}
	\label{table:cen-xray-ratios}
	\begin{tabular}{|c|c|c|c|c|c|}
		\hline
		\hline
		Label						&
		$\lambda$ / \AA					&
		$L / L$(\emlineforb{Fe}{x})			&
		$\Gamma(T) / \Gamma(T)$(\emlineforb{Fe}{x})	&
		$\dot{M} / \dot{M}$(\emlineforb{Fe}{x})		&
		$\dot{M}$(\emlineforb{Fe}{x}) / \Mdotunit{}	\\
		\hline
		\emline{Fe}{xvii}	&	15.01	& 354		& 359 	& 0.99	& 1.6	\\
		\emline{Fe}{xvii}	&	15.26	& 103		&  97 	& 1.06	& 1.5	\\
		\emline{Fe}{xvii}	&	16.78	& 123		& 180 	& 0.68	& 2.5	\\
		\emline{Fe}{xvii}	&	17.05	& 551		& 224 	& 2.46	& 1.2	\\
		\emline{N}{vii}		&	24.78	& 329		& 146 	& 2.25	& 4.3	\\
		\hline
	\end{tabular}
\end{table*}
%%%%%%%%%%%%%%%%%%%%%%%%%%%%%%%%%%%%%%%%%%%%%%%%%%%%%%%%%%%%%%%%%%%%%%%%%%%%%%%%%

\section{Coronal Gas as a Cooling Condensation}\label{sec:cooling-condens}

\par
Deciphering the lack of any \emlineforb{Fe}{xiv} emission is
fundamental to our understanding of the state of the coronal
gas.
If the coronal gas is indeed cooling out of the ambient
ICM, the mass deposition rate through \emlineforb{Fe}{xiv}
should be consistent with the X-ray and coronal line rates.
We now leverage our cooling models to convert the upper limit
to the luminosity ratio to a limit for the ratio of mass
deposition rates.

\par
Table~\ref{table:gamma-rates} lists our estimates for the $\Gamma(T)$ ratio
between each of the undetected lines and the detected \emlineforb{Fe}{x}.
They amount to the luminosity ratios expected for a simple cooling
flow model of uniform deposition rate.
Interestingly, the ratio of $\Gamma(T)$ functions for \emlineforb{Fe}{xiv}
is greater than the observed limit, suggesting that if the coronal gas was
cooling freely from the hot ICM, the \emlineforb{Fe}{xiv} would probably
have been observed.

\subsection{Continuum Subtraction}\label{sec:continuum}

\par
The C11a upper limits depend sensitively on the continuum subtraction
method employed (see Sec.{} 3.3 therein).
In particular, single stellar population model spectra are
dominated by stellar features below 6000\unit{\AA}, which
vary on scales comparable to the expected coronal line widths,
complicating the analysis.
C11a stress that the uncertainties in the stellar continuum
subtraction may introduce a larger error into the reported upper
limit for \emlineforb{Fe}{xiv}.

\par
We note that our calculations yield a $\Gamma(T)$ function
ratio of 2.85 for \emlineforb{Fe}{xiv} over \emlineforb{Fe}{x}.
By contrast, the older models of \citet{GraneySarazin1990}
predict values less than unity.
This is a consequence of the changes in the \emlineforb{Fe}{xiv}
$\lambda$5303\unit{\AA} emissivity discussed in Sec.~\ref{sec:comp-graney}.
In other words, our findings further exacerbate the non-detection
of \emlineforb{Fe}{xiv}.

\par
While we acknowledge the complications involved in estimating accurate fluxes
for the \emlineforb{Fe}{xiv} line, in the following, we assume that the
non-detection is not due to such technicalities, and explore its consequences
for the state of the system.

\subsection{Mass Deposition Rates}\label{sec:mass-rates}

\par
Accounting for the upper limits of C11a, the $\Gamma(T)$ ratios can
be converted to upper limits for the mass deposition rate through
\begin{equation}
	\frac{\dot{M}_\mathrm{line}}{\dot{M}_\mathrm{[Fe \; X]}} <
	\frac{L_\mathrm{line}}{L_\mathrm{[Fe \; X]}}		\, 
	\frac{\Gamma_\mathrm{[Fe \; X]}}{\Gamma_\mathrm{line}}	\,	.
	\label{eqn:mass-rate-limit}
\end{equation}
The maximum rates are presented in Table~\ref{table:mass-rates}.
Our calculations suggest that the mass deposition rate
at a million K (\emlineforb{Fe}{x}) is twice as high
as the rate at two million K (\emlineforb{Fe}{xiv}),
or that more gas is cooling from a million K, than is
cooling to that temperature.
This contradicts cooling flow observations, which suggest
that the mass deposition rate is either constant or increases
with temperature.

\par
The lack of any significant \emlineforb{Fe}{xiv} emission
suggests that the coronal gas cannot have formed in situ
as a cooling condensation of the surrounding ICM.
It is more likely that, as the X-ray morphology suggests, the gas
originated closer to the galactic nucleus, and it has been displaced
to its current position by a dynamical process, such as uplifting
by radio bubbles, and possibly heated.
Other possibilities, such as conduction and mixing, are discussed
in Section~\ref{sec:discussion}.

%%%%%%%%%%%%%%%%%%%%%%%%%%%%%%%%%%%%%%%%%%%%%%%%%%%%%%%%%%%%%%%%%%%%%%%%%%%%%%%%%
%				TABLE						%
%%%%%%%%%%%%%%%%%%%%%%%%%%%%%%%%%%%%%%%%%%%%%%%%%%%%%%%%%%%%%%%%%%%%%%%%%%%%%%%%%
\begin{table}
	\centering
	\caption
	{
		Ratio of gamma functions, $\Gamma_\mathrm{line} / \Gamma_\mathrm{[Fe \, X]}$,
		for cooling calculations of the five-component cooling flow model of
		\citet{Sanders+2008-Cen}.
		The first two columns identify the spectral line, while
		the remaining columns list results for NEI and CIE cooling under
		isochoric (CD) and isobaric (CP) conditions.
	}
	\label{table:gamma-rates}
	\begin{tabular}{|l|c|c|c|c|c|}
		\hline
		\hline
		\multicolumn{2}{c}{Spectral Line}	\\
		Label		& $\lambda$ / \AA	&NEI--CP& CIE--CP	& NEI--CD	& CIE--CD	\\
		\hline
		\emlineforb{Fe}{xiv}	& 5303			& 2.84	&	2.84	&	2.85	&	2.87	\\
		\emlineforb{Ca}{xv}	& 5445			& 0.24	&	0.25	&	0.24	&	0.25	\\
		\emlineforb{Ca}{xv}	& 5695			& 0.56	&	0.57	&	0.56	&	0.58	\\
		\emlineforb{Ni}{xv}	& 6700			& 0.07	&	0.07	&	0.07	&	0.07	\\
		\hline
	\end{tabular}
\end{table}
%%%%%%%%%%%%%%%%%%%%%%%%%%%%%%%%%%%%%%%%%%%%%%%%%%%%%%%%%%%%%%%%%%%%%%%%%%%%%%%%%
\begin{table}
	\centering
	\caption
	{
		Upper limits to the ratios of mass deposition rates, obtained with
		the results of Table~\ref{table:gamma-rates}, and eqn.~(\ref{eqn:mass-rate-limit}).
		The meaning of the columns is the same as in Table~\ref{table:gamma-rates}.
	}
	\label{table:mass-rates}
	\begin{tabular}{|l|c|c|c|c|c|}
		\hline
		\hline
		\multicolumn{2}{c}{Spectral Line}	\\
		Label			& $\lambda$ / \AA	&NEI--CP& CIE--CP	& NEI--CD	& CIE--CD	\\
		\hline
		\emlineforb{Fe}{xiv}	& 5303			& 0.45	&	0.45	&	0.45	&	0.45	\\
		\emlineforb{Ca}{xv}	& 5445			& 1.19	&	1.17	&	1.20	&	1.16	\\
		\emlineforb{Ca}{xv}	& 5695			& 1.94	&	1.90	&	1.96	&	1.89	\\
		\emlineforb{Ni}{xv}	& 6700			& 4.34	&	4.28	&	4.38	&	4.26	\\
		\hline
	\end{tabular}
\end{table}
%%%%%%%%%%%%%%%%%%%%%%%%%%%%%%%%%%%%%%%%%%%%%%%%%%%%%%%%%%%%%%%%%%%%%%%%%%%%%%%%%

\subsection{Differential Extinction}\label{sec:extinction}

\par
Before we investigate the properties of the coronal gas in more
detail, we explore the possibility that the lack of \emlineforb{Fe}{xiv}
emission is due to extinction.
Having established that the detected coronal line is optically thin,
steep extinction laws are required to account for the observations.

\par
\begin{figure}
	\begin{centering}
	\includegraphics{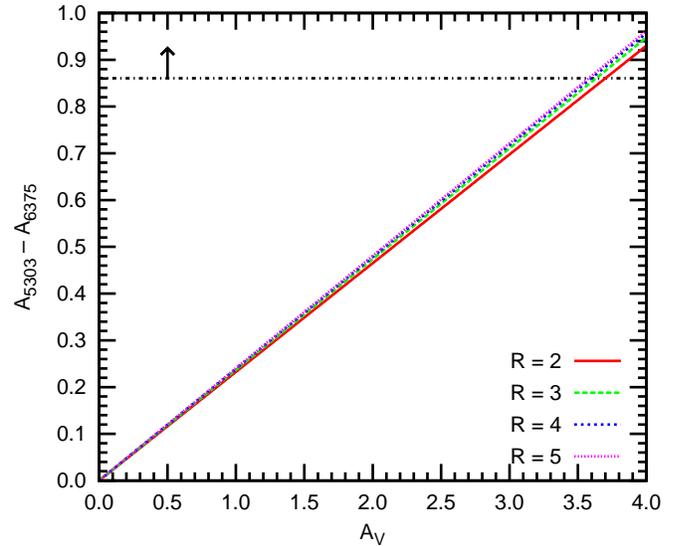}
	\end{centering}
	\caption[Extinction of Line Ratio Dependence on $A_V$.]
	{
		Dependence of differential extinction on the $V$-band
		extinction for a number of different $R$ extinction laws.
		The dot-dashed line shows the limit derived from
		the observations for a $\Gamma(T)$ ratio of 2.85
		(see Table~\ref{table:mass-rates}).
	}
	\label{fig:extinction}
\end{figure}

\citet{Mittal+2011-cen-herschel} reported the detection of extended
dust emission in NGC 4696.
If the X-ray cool gas resides on the far side of the radio source,
extinction may account for the non-detection of \emlineforb{Fe}{xiv}.
Assuming a constant mass deposition rate in the 1--2~million K
range, the reddening between \emlineforb{Fe}{xiv} and \emlineforb{Fe}{x} is
\begin{equation}
	A_{5303}-A_{6375} > -2.5\, \log_{10}\Biggl(
		\frac{L_\mathrm{[Fe \; XIV]}}{L_\mathrm{[Fe \; X]}}			\, 
		\frac{\Gamma_\mathrm{[Fe \; X]}}{\Gamma_\mathrm{[Fe \; XIV]}}	\Biggr)	\,	.
	\label{eqn:color-excess}
\end{equation}
Because these wavelengths lie in the linear part of the extinction curve
\citep{Fitzpatrick1999}, they are largely independent of the reddening
constant, $R = A_V / E(B-V)$, and may be approximated by
$A_{5303}-A_{6375} \simeq 0.24 \, A_V$.
For the values shown in Tables~\ref{table:cen-flux-ratios}
and \ref{table:mass-rates}, we obtain $A_V > 3.6125$.
The detailed dependence of the differential extinction
on the $V$-band extinction is shown in Figure~\ref{fig:extinction}.
Note that mass deposition rates that decrease with declining temperature,
as suggested by X-ray spectroscopy \citep{Sanders+2008-Cen},
would require even heavier extinction.

\par
The required minimum extinction is difficult to reconcile with observations,
as it exceeds the maximum $V$-band extinction estimates in the vicinity of
NGC~4696.
\citet{Sparks+1989-cen} reported a maximum of 0.4 at the south-western end
of the dust lane, based on $V$- and $R$-band imaging.
On the other hand, \citet{Farage+2010-Cen-IFU} estimated a maximum of
$\sim$0.7, 3\arcsec{} north-west of the nucleus, from the observed
$\ha / \hb$ ratio, and the standard assumption for active galaxies
that its intrinsic value is higher than the Case B value, at 3.1.
\citet{Canning+2011-cen} reported maximum extinctions of
$A_V$$\sim$2--2.5 (see fig.{} B1 in their appendix) for a
Case B $\ha / \hb$ ratio.
For values appropriate to active galaxies, however, they obtained
results consistent with \citet{Farage+2010-Cen-IFU}.
Similar results have been obtained by other authors
\citep[e.g., see references in][]{Canning+2011-cen}.

\par
In any case, there is not significant overlap between the gas cloud
of interest (``box 1'' in C11a) and regions of high extinction according
to fig.~B1 of \citet{Canning+2011-cen}, although C11a point out that
assessing the connection between the coronal line gas, and the optical
filaments and dust lane will require deeper observations.

%%%%%%%%%%%%%%%%%%%%%%%%%%%%%%%%%%%%%%%%%%%%%%%%%%%%%%%%%%%%%%%%%%%%%%%%%%%%%%%%%
%				FIGURE						%
%%%%%%%%%%%%%%%%%%%%%%%%%%%%%%%%%%%%%%%%%%%%%%%%%%%%%%%%%%%%%%%%%%%%%%%%%%%%%%%%%
%%%%%%%%%%%%%%%%%%%%%%%%%%%%%%%%%%%%%%%%%%%%%%%%%%%%%%%%%%%%%%%%%%%%%%%%%%%%%%%%%

\par
Also note that these conclusions are in qualitative agreement with
the modelling of \citet{Werner+2013-Virgo}.
These authors postulated the presence of intervening cold absorbers
intermixed with the X-ray filaments in Virgo, and found that a column
density of 1.6$\times$10$^{21}$\unit{cm^{-2}} may account for the lack
of X-ray emission below 0.5\unit{keV}.
If a column density of that magnitude occurs in Centaurus as well,
it would translate to reddening $E(B-V) = 0.276$, according to
the dust-to-gas relation of \citet{Bohlin+1978}, namely,
$N(\hi{} + \htwo{}) = 5 \times 10^{21} E(B-V)$.
However, such a column can only account for the preferential
extinction of the \emlineforb{Fe}{xiv}, if the reddening constant
has a value $R \gtrsim 13$.
We can therefore rule out this possibility, as well.

\par
We conclude that extinction cannot account for
the lack of \emlineforb{Fe}{xiv} emission.

\subsection{Heating}\label{sec:heating}

\par
Heating may prevent a gas from cooling to low temperatures
either by establishing a temperature floor, or by increasing
the thermal energy of the gas and reducing its net cooling rate.
Because the details of how heating affects gas at different
temperatures are uncertain, simulations of the latter scenario fall
beyond the scope of the present paper.
In this paper, we assume that heating, if it has occurred, has merely
established a temperature floor.
If the floor falls in the temperature range probed by \emlineforb{Fe}{x},
it will affect the deduced mass deposition rates.

\par
\begin{figure*}
	\begin{centering}
	\includegraphics{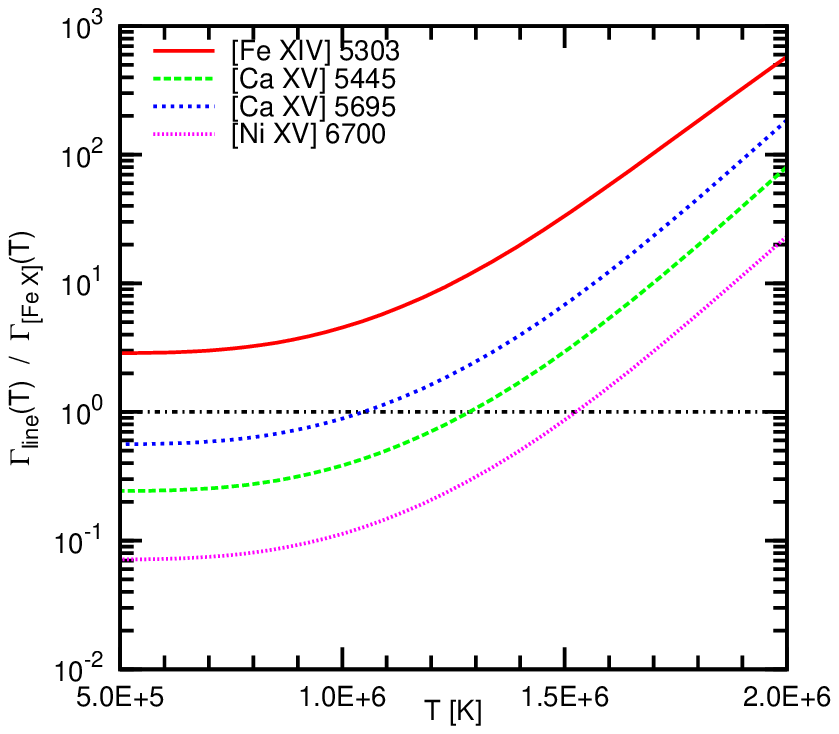}
	\includegraphics{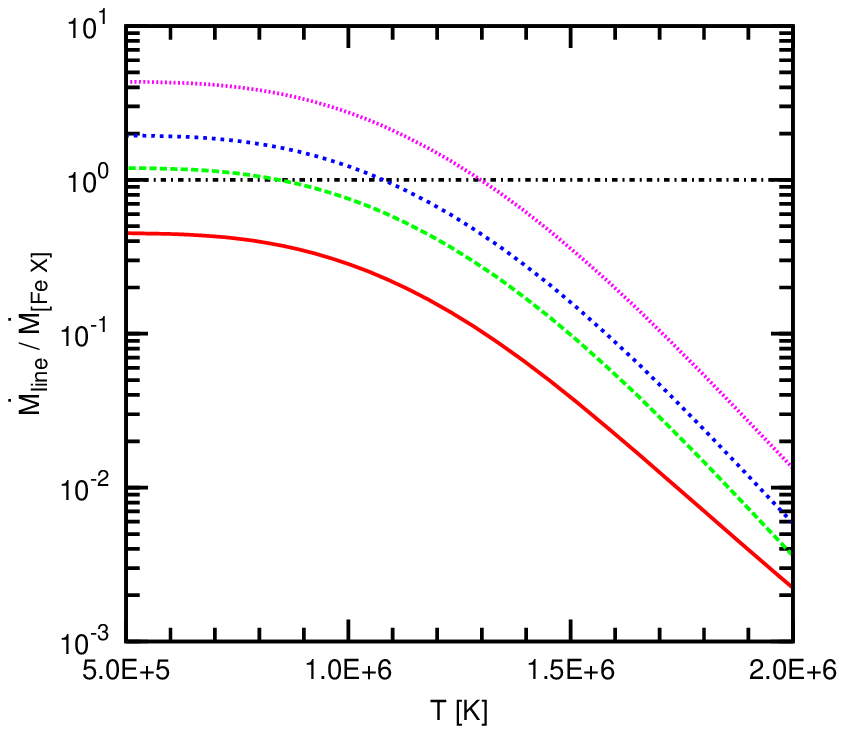}
	\end{centering}
	\caption[Cooling flow model line ratios with temperature.]
	{
		(Left panel) Ratio of $\Gamma(T, \, T_\mathrm{max})$
		for each of the non-detected lines against the detected
		\emlineforb{Fe}{x}, as a function of temperature.
		(Right panel) Maximum deposition mass rates for each of
		these lines, computed as the r.h.s.{} of
		eqn.~(\ref{eqn:mass-rate-limit}), as a function of
		temperature.
	}
	\label{fig:cool-flow-mass-rates}
\end{figure*}

Figure~\ref{fig:cool-flow-mass-rates} explores the temperature dependence
of the computed gamma ratios, as well as the maximum deposition rate ratios
subject to the observed luminosity upper limits.
The $\Gamma(T)$ functions are computed between the initial temperature
in the simulation and the temperature floor in question, $T$.
In practice, an emission line contributes significantly over a limited
temperature range.
The range for the plots is chosen to cover most of the range probed
by the \emlineforb{Fe}{x} line (see Fig.~\ref{fig:graney90-comp}).
At the left end of these plots, the $\Gamma(T)$ and $\dot{M}$ ratios are near
the asymptote values shown in Table~\ref{table:mass-rates}.
At higher temperatures, the amount of cooling due to \emlineforb{Fe}{x} diminishes
causing the $\Gamma(T)$ ratios to increase, and the mass deposition rate ratios
to decrease.

%%%%%%%%%%%%%%%%%%%%%%%%%%%%%%%%%%%%%%%%%%%%%%%%%%%%%%%%%%%%%%%%%%%%%%%%%%%%%%%%%
%				FIGURE						%
%%%%%%%%%%%%%%%%%%%%%%%%%%%%%%%%%%%%%%%%%%%%%%%%%%%%%%%%%%%%%%%%%%%%%%%%%%%%%%%%%
%%%%%%%%%%%%%%%%%%%%%%%%%%%%%%%%%%%%%%%%%%%%%%%%%%%%%%%%%%%%%%%%%%%%%%%%%%%%%%%%%

\par
It is important to notice that the $\Gamma(T)$ ratio for \emlineforb{Fe}{xiv}
is always higher than unity, suggesting that cooling due to this line
is more important than cooling due to the \emlineforb{Fe}{x} {\em over the
entire temperature range.}
This is partly because the gas cools faster through the temperature range
probed by \emlineforb{Fe}{x}.

\par
The right panel of Fig.~\ref{fig:cool-flow-mass-rates} illustrates that the
mass deposition rate consistent with the observed upper limits for \emlineforb{Fe}{xiv}
is {\em always} lower than the rate traced by \emlineforb{Fe}{x}.
In fact, the rate becomes progressively lower at higher temperatures, as the
cooling due to \emlineforb{Fe}{x} diminishes.
If extinction is at work, even higher values
than previously computed are required to account
for the increasing $\Gamma(T)$ ratios.

\par
These arguments show that a simple temperature floor cannot account
for the lack of emission from gas hotter than $\sim$million K in the
observations of C11a, and support our conclusion that the coronal gas
cannot be a cooling condensation of the surrounding ICM.

\section{Constraints from \emlineforb{F\MakeLowercase{e}}{xiv}}\label{sec:FeXIV-constraints}

\par
As discussed in Section~\ref{sec:properties}, constraining
the properties of the coronal gas requires an estimate of
the volume it occupies.
The coronal gas was detected along the cool X-ray plume
\citep{SandersFabian2002}, and most likely it is embedded
within the X-ray filament.
However, the connection between the X-ray and coronal phases,
as well as the optical filaments, remains unclear (C11a).

\par
In this section we consider the possibility that the coronal
gas is projected on to the X-ray cool filament by chance, and
arrive at an estimate for its volume fraction.
For this purpose, we assume that gas cooling may be described
by the cooling flow formalism, in the sense that line
luminosities are given by eqn.~(\ref{eqn:cflow-lum}).
Because cooling is thought to occur ``locally'',
we take the mass deposition rate to be constant.
Differences in the cooling rate are encapsulated
in the $\Gamma(T)$ function, and fully accounted
for when taking line ratios.
We also adopt the abundances of the 5$\times$VAPEC model
fits of \citet{Sanders+2008-Cen}, shown in the last column of
Table~\ref{table:CentaurusAbundances-2008}.
Note, however, that the differences to the abundances appropriate to the
five component cooling flow model are within the statistical errors (10\%)
for most elements.

\subsection{Temperature Ceiling}\label{sec:temp-ceiling}

\par
\begin{figure}
	\begin{centering}
	\includegraphics{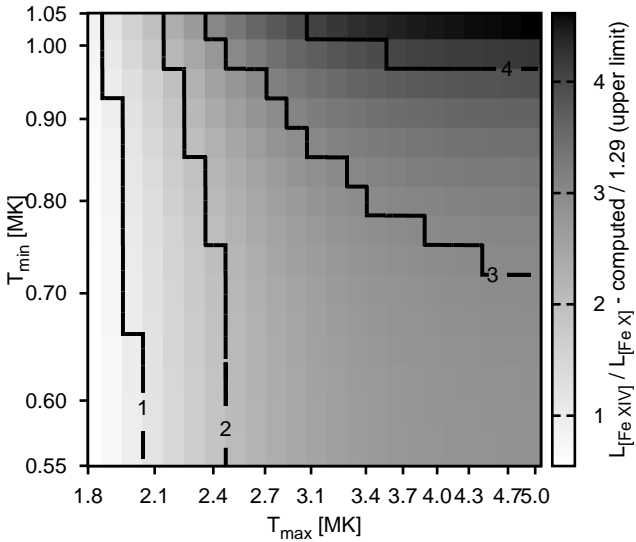}
	\end{centering}
	\caption[Luminosity ratio calculations.]
	{
		Ratio of the computed luminosity ratio of
		\emlineforb{Fe}{xiv} over \emlineforb{Fe}{x}
		in units of the reported upper limit (1.29).
		The results shown are for isobaric cooling.
		The temperature ceiling varies along the x-axis, while the
		temperature floor varies along the y-axis.
		The upper limit is consistent with a
		temperature ceiling of $\sim$2.1 million K.
	}
	\label{fig:lratio-temp}
\end{figure}

We consider the luminosity ratio between the \emlineforb{Fe}{xiv}
and \emlineforb{Fe}{x} lines computed over the temperature
intervals which contribute 99\% of the respective total line
luminosity.
Our goal is to obtain an estimate for the maximum temperature
of the coronal gas that is consistent with the reported upper
limit (\emlineforb{Fe}{xiv} / \emlineforb{Fe}{x} $< 1.29$).
We let the maximum temperature of the gas vary between the temperature
at which the cooling efficiency of \emlineforb{Fe}{xiv} peaks (1.8 million K)
and the upper end of the line temperature range.
We also explore the robustness of these calculations in the presence
of a temperature floor, which we simulate by letting the minimum
temperature probed by \emlineforb{Fe}{x} be at most equal to the
temperature of peak cooling efficiency ($\sim$1 million K) and
no less than the lower end of the line temperature range.
(The temperature range probed by \emlineforb{Fe}{x} is
$T^X_\mathrm{min} \le T \le T^X_\mathrm{max}$.)
Figure~\ref{fig:lratio-temp} shows that the computed luminosity ratio
is consistent with the reported upper limit for temperatures no greater
than $2.1$~million K.
This is true for both isochoric and isobaric cooling.
The presence of a temperature floor does not modify these results
significantly.
In the following, we adopt a conservative temperature ceiling of 2.1~million K.

%%%%%%%%%%%%%%%%%%%%%%%%%%%%%%%%%%%%%%%%%%%%%%%%%%%%%%%%%%%%%%%%%%%%%%%%%%%%%%%%%
%				FIGURE						%
%%%%%%%%%%%%%%%%%%%%%%%%%%%%%%%%%%%%%%%%%%%%%%%%%%%%%%%%%%%%%%%%%%%%%%%%%%%%%%%%%
%%%%%%%%%%%%%%%%%%%%%%%%%%%%%%%%%%%%%%%%%%%%%%%%%%%%%%%%%%%%%%%%%%%%%%%%%%%%%%%%%

\subsection{Volume Filling Factors}\label{sec:volume-filling}

\par
The volume occupied by gas in the temperature range $(T, T+dT)$ is
$dV = \dot{M} \, dt / \rho$.
For a constant mass deposition rate, the volume of gas in $(T_1, T_2)$ is
\begin{equation}
	V	= \dot{M}	\,	\int^{T_2}_{T_1} dt / \rho 
		= \dot{M}	\,	Z(T_1, T_2)	\,	.
	\label{eqn:vol-time}
\end{equation}
The total volume of the cooling gas may be obtained by extending
the integral limits to include the entire range of temperatures
covered by the gas.
Then, the volume fraction for gas over a limited temperature range
is given by the ratio
\begin{equation}
	f_V = Z(T_1, T_2) \Bigl/ Z(T_\mathrm{min}, T_\mathrm{max})	\,	.
	\label{eqn:vol-fraction}
\end{equation}

\par
\begin{figure}
	\begin{centering}
	\includegraphics{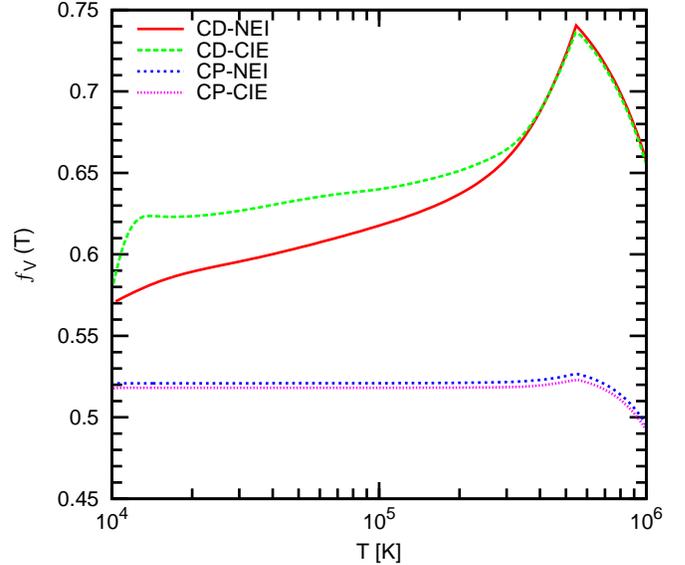}
	\end{centering}
	\caption[Volume Fraction for \emlineforb{Fe}{x} phase.]
	{
		Volume fraction of the phase that emits \emlineforb{Fe}{x}
		as a function of the temperature floor.
		The calculations are done for temperatures below the cooling
		efficiency peak. 
		The line cooling efficiency effectively drops to zero below
		0.55 million K.
	}
	\label{fig:vol-fraction}
\end{figure}

We compute the volume fraction of the \emlineforb{Fe}{x} phase as a function
of the temperature floor, which we let extend to 10,000\unit{K}.
The integration in the numerator of eqn.~(\ref{eqn:vol-fraction})
is carried out over the range $(T, T^X_\mathrm{max})$, where
$T \ge T^X_\mathrm{min}$.
In the denominator the integration is carried out over $(T, T_\mathrm{max})$,
where $T_\mathrm{max}$ is the adopted temperature ceiling.
As Figure~\ref{fig:vol-fraction} shows, typical values are in the range of 50--75\%.
For isobaric cooling, the volume fraction obtains an asymptotic value at low
temperatures, due to compression.
For isochoric cooling, by contrast, the volume fraction declines with decreasing
temperature below 200,000\unit{K}.
The NEI calculations predict a slightly lower volume fraction.
This is due to reduced cooling relative to CIE \citep{GnatSternberg07},
which leads to gas surviving for longer at such temperatures.

%%%%%%%%%%%%%%%%%%%%%%%%%%%%%%%%%%%%%%%%%%%%%%%%%%%%%%%%%%%%%%%%%%%%%%%%%%%%%%%%%
%				FIGURE						%
%%%%%%%%%%%%%%%%%%%%%%%%%%%%%%%%%%%%%%%%%%%%%%%%%%%%%%%%%%%%%%%%%%%%%%%%%%%%%%%%%
%%%%%%%%%%%%%%%%%%%%%%%%%%%%%%%%%%%%%%%%%%%%%%%%%%%%%%%%%%%%%%%%%%%%%%%%%%%%%%%%%

\section{Coronal Gas Properties}\label{sec:properties}

\par
The \emlineforb{Fe}{x} luminosity may be used to infer the coronal gas properties.
Motivated by the optical and X-ray filaments in the same region, we approximate
the coronal gas as a cylinder, and take its diameter and height equal to the size
reported by C11a, i.e., $2R \sim 700$\unit{pc}.
Below we first explore the scenario that the coronal gas is independent of
the X-ray plume, and then that it is embedded within the plume.

\subsection{Independent Structure}\label{sec:properties:independent}

\par
We assume that the temperature floor lies below the range of \emlineforb{Fe}{x}
emission, but not by much.
This allows us to adopt a volume filling factor of 60\%.
We accumulate line fluxes according to eqns.
(\ref{eqn:flux-integral})--(\ref{eqn:flux-integral:isobaric}),
and solve the standard equation
\begin{equation}
	L^X	=	f_V \, V \, n_e \, n_H \, \Lambda^X(T)	\,	,
	\label{eqn:luminosity}
\end{equation}
for the hydrogen density, $n_H$.
The computed accumulated fluxes are rescaled by the square of the density
at the temperature of peak \emlineforb{Fe}{x} cooling, $\sim$million K, for
that purpose.

\par
All cooling runs are consistent with a hydrogen density of $\sim$0.2\unit{cm^{-3}}.
This includes isochoric and isobaric calculations, in and out of collisional
equilibrium.
The obtained density is not sensitive to the adopted volume fraction,
$n_H \propto f_V^{-1/2}$, and values in the range 50--75\% lead to small
density variations.
The coronal gas mass is about 1.5$\times$10$^6$\unit{M_\odot}.

\par
If we assume that the coronal gas is thermally supported, pressure equilibrium
translates to a distance of the cloud to the radio source of about 20\unit{kpc}.
However, C11a concluded that the \emlineforb{Fe}{x} is turbulently broadened at
$\sim$300\unit{km \, s^{-1}}.
We have therefore assumed that transonic turbulence ($\upsilon_\mathrm{turb} \sim c_s$, 
where $\upsilon_\mathrm{turb}$ is the turbulent velocity, and $c_s$ is the sound speed)
contributes to the pressure budget of the coronal cloud.
This leads to a linewidth of $\sim$250\unit{km \, s^{-1}}, in decent agreement
with the observations.
Pressure equilibrium, then, translates to a distance of $\sim$8\unit{kpc},
suggesting that the X-ray arc is located within NGC~4696.
The projected distance from the nucleus is $\sim$3\unit{kpc},
so the line-of-sight distance to the nucleus is $\sim$7.5\unit{kpc}.

\par
The sound crossing time of the coronal cloud is about 4.5\unit{Myr}.
By contrast, the gas cooling time is $\sim$0.6\unit{Myr} for isochoric
cooling, and $\sim$0.8\unit{Myr} for isobaric cooling.
In either case, the cooling time is a small fraction of the crossing time,
which suggests that the gas cooling proceeds under conditions of constant
density (isochoric).

\par
In cooling flow models, the gas is cooling such that its temperature
follows the underlying gravitational potential \citep{McNamaraNulsen2007}.
This property allows to test for the possibility that the gas has been
displaced to the observed location.
An alternative method is to compare the entropy content of the coronal
gas to that of the surrounding ICM, as described by the fitting forms
of \citet{Graham2006-CentaurusCluster}.
Note that the ICM follows the familiar $K(r) \propto r^{0.96}$ entropy law
down to $\sim$1\unit{kpc} from the nucleus.
We find that the coronal gas carries about $1/90$ of the entropy in the
surrounding X-ray gas.
If the entropy distribution holds at sub-kpc scales, the coronal gas
entropy is consistent with a distance of $\sim$0.1\unit{kpc} from the
nucleus, and therefore consistent with a displacement scenario.

\subsection{Embedded In X-ray Plume}\label{sec:properties:embedded}

\par
If the coronal gas is embedded within the X-ray plume, we can require
that it be in pressure equilibrium with the hot phase, and we can use
the coronal luminosity to constrain its volume fraction.

\par
The X-ray gas is estimated to have a density of 0.13\unit{cm^{-3}},
and a temperature of 0.8\unit{keV} \citep{SandersFabian2002}.
We assume that the coronal gas covers the range of temperatures
probed by \emlineforb{Fe}{x}.
A wider range would introduce a 50-75\% correction to its volume
fraction, as discussed previously.
We also assume that it is supported by transonic turbulence
($\upsilon_\mathrm{turb} \sim c_s$, where
$\upsilon_\mathrm{turb}$ is the turbulent velocity, and
$c_s$ is the sound speed).
The condition of pressure equilibrium then gives the hydrogen
density of the coronal gas as
\begin{equation}
	n_H =	\frac{n_H^\mathrm{ICM}}{1 + \gamma / 3 }	\,
		\frac{T^\mathrm{ICM}}{T}			\,	.
	\label{eqn:pressure-sonic-turbulence}
\end{equation}
Numerically, the density is 0.74\unit{cm^{-3}}.
Then, eqn.~(\ref{eqn:luminosity}) yields $f_V \sim 6$--7\%
for isochoric and isobaric cooling, respectively.
Assuming that the coronal gas forms one coherent cylindrical filament,
its radius will be $\sqrt{f_V} \, R$, or $\sim$85--95\unit{pc}.
The sound crossing time is 1--1.2\unit{Myr}, while the cooling
time is 0.2--0.3\unit{Myr}.
In this case too, then, the gas is cooling isochorically.
Its mass is now $\sim$5$\times$10$^5${$\thinspace M_\odot$},
and its entropy is $\sim$30 times lower than the cool X-ray gas.

\section{Discussion}\label{sec:discussion}

\par
Our analysis indicates that the coronal gas is a young dynamical
feature in NGC~4696. 
It is unlikely that it has cooled out of the surrounding ICM,
or \emlineforb{Fe}{xiv} would have been observed.
Yet, it coincides with the X-ray cool filament in projection,
which suggests that the two phases may in fact be co-spatial.
The coincidence with the optical filaments is more tenuous
and requires further investigation (C11a).

\par
If the optical and coronal phases are not related to each other,
the coronal gas could possibly have originated as warm ($\sim$10$^4$\unit{K})
gas that was vigorously heated, and in which cooling has recently
been re-established.
The energy required to heat the coronal gas from 10$^4$\unit{K}
to 10$^6$\unit{K} is 4$\times$10$^{53}$\unit{erg}.
According to \citet{Taylor+2006-centaurus-radio-source}, the power
of the active galactic nucleus (AGN) of NGC~4696 is currently
$< 10^{40}$\unit{erg \, s^{-1}}, which suggests that the AGN
outburst lasted at least 1.5~million years.
These calculations assume 100\% efficiency and do not account for the
work required to lift the gas out to 9\unit{kpc}, so likely the outburst
lasted longer and/or was more powerful.
The time-scale is plausible, as it is comparable with the
$\sim$2.5\unit{Myr} periodicity of the AGN in M87 \citep{Nulsen+2007}.
Higher final temperatures ($> 10^6$\unit{K}) require
higher outburst powers, durations, or both.
However, this scenario is not satisfactory, because it requires
some fine-tuning in order to account for the fact that we get to
observe freely cooling gas of such short cooling time.

\par
A more natural explanation would involve the coronal gas
as a phase of a steady-state system.
Most likely the optical, coronal, and X-ray phases are related
to each other.
Then, the coronal gas may be the conductive or mixing interface
between the optical, 10$^4$\unit{K}, filaments and the cool X-ray
gas, 10$^7$\unit{K}.
Examples of conductive interfaces in rough agreement with the
coronal gas properties may be found in
\citet{BoehringerFabian1989-conduction}.
On the other hand, \citet{BegelmanFabian1990-mixing-layers}
present crude estimates on the structure of mixing layers.
The temperature distribution of the gas in the mixing layer
is analytically unknown.
The geometric mean temperature is around 300,000\unit{K}, subject
to unknown, order of unity efficiency factors.
This is effectively consistent, to within a factor of a few,
with the temperature of the \emlineforb{Fe}{x} gas.
In addition, assuming that the ambient ICM is turbulent, and that
the turbulent velocity is close to the sonic speed, we obtain
from their equation (2) that the depth of the mixing layer is
roughly 1.3\unit{kpc}, consistent with the X-ray filament width
to a factor of 2.
Based on the sharpness of the optical filaments
\citep{Crawford2005-cen}, the ICM is thought to not be
particularly turbulent, which should improve the agreement
with the observations.
In fact, plugging in that equation the previous estimate of
$\sim$90\unit{pc} for the thickness of the mixing layer, we
obtain highly subsonic turbulent motions in the plume, such
that ${\cal M} < 0.1$.

\par
Note that in the simulations of \citet{Esquivel+2006},
which do not parametrize uncertainties with efficiency
factors, the mixing layer that forms between the $10^4$\unit{K}
and $10^7$\unit{K} phases has a density-weighted temperature
of 1--2 million K.
However, these simulations do not reach steady-state after
3\unit{Myr} of integration, and do not include the effects
of NEI.
The more recent simulations of \citet{KwakShelton2010-NEI-mixing}
do include non-equilibrium effects, albeit in an idealized
two-dimensional, unmagnetized setting.
They find that in the simulation most similar to our case
(Model F, which has a hot phase temperature of
3$\times$10$^6$\unit{K}) the mixing layer does not reach
steady-state by the end of their runs (80 Myr), unlike
simulations at lower temperatures for the hot phase.
However, they point out that the rate of growth
for the mixing layer correlates with the shear speed.
This suggests that higher shear speeds may be required
to produce a steady-state mixing layer in the X-ray plume.

\par
\begin{figure*}
	\begin{centering}
	\includegraphics{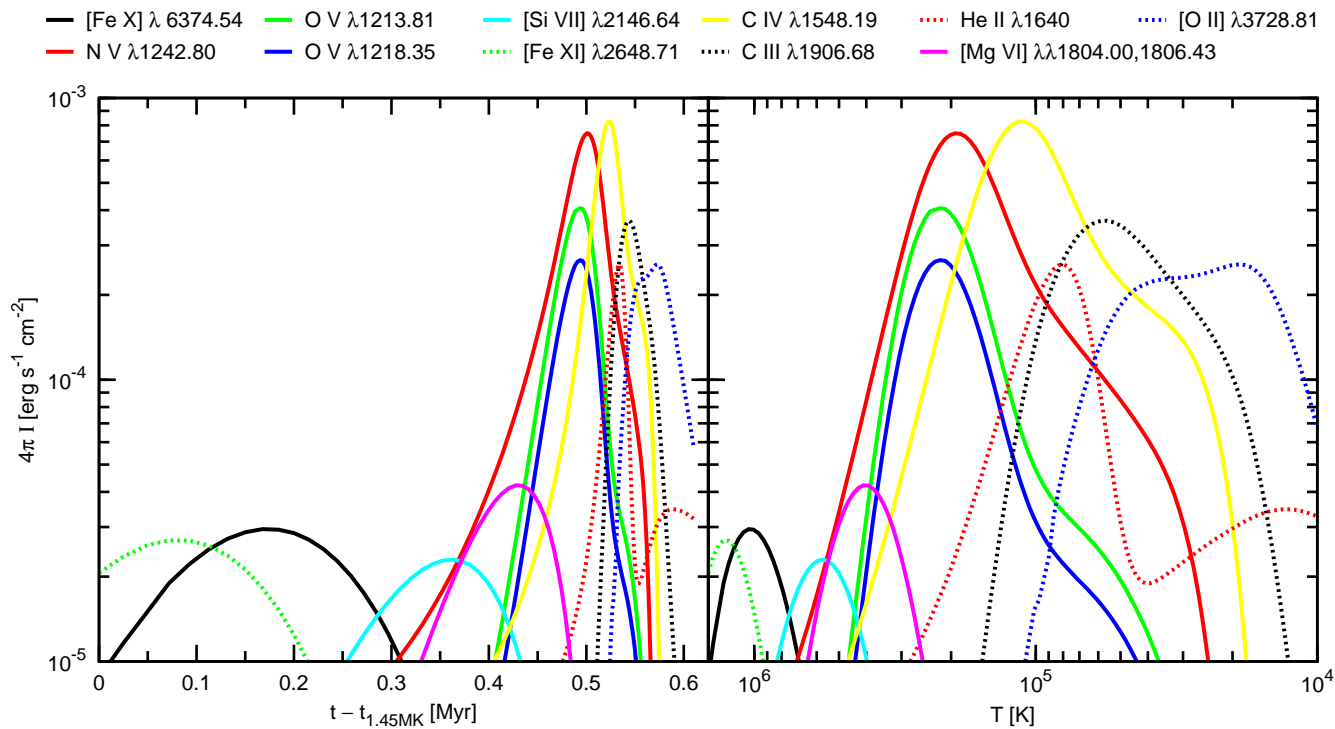}
	\end{centering}
	\caption[Line emission with time and temperature.]
	{
		Intensity of various lines produced by a gas column as a function
		of time (left) and decreasing temperature (right) for our cooling
		calculations.
		The column length is equal to the coronal cloud diameter, and the
		hydrogen density is set to 0.2\unit{cm^{-3}}.
		Time is measured since the moment when the temperature is 1.45~million
		K.
	}
	\label{fig:line-emiss}
\end{figure*}

Signatures of gas at intermediate temperatures may help resolve
the state and origin of the coronal gas.
Obviously, conduction and mixing models involve multi-temperature
gas distributions, so detection of emission lines from intermediate
temperature may be expected.
Such emission lines have been observed previously in Virgo,
where \emline{C}{iv} $\lambda$1549 due to $10^5$\unit{K}
has been reported by \citet{Sparks+2009-Virgo-CIV,Sparks+2012-Virgo-CIV}.
Or, in Abell 426, and Abell 1795, where gas at $10^{5.5}$\unit{K}
gives rise to \emline{O}{vi} $\lambda\lambda$ 1032, 1038\unit{\AA}
\citep{Bregman+2006}.
In Figure~\ref{fig:line-emiss} we present the best
candidate lines identified in our cooling simulations.
These lines are brighter than the detected \emlineforb{Fe}{x},
also shown.
The evolution with time and temperature indicate that some of
these lines although very bright are associated with gas of
short cooling time, and may not be easily detected.
However, if the coronal gas is cooling in a steady state,
some of these lines will likely be observable in the ultraviolet.

\par
The non-detection of either \emlineforb{Fe}{x} or
\emlineforb{Fe}{xiv} in the north west end of the
X-ray arc poses difficulties to the mixing and conduction
scenarios.
The north-west end of the arc lies close to the dust lane,
so absorption is a likely cause for the non-detections,
without requiring fine tuning of the obscuring column density.
Alternatively, or in addition, if that gas lies deeper into
the potential well, the mean temperature of the mixing layer
may be lower than that at south east end, which would diminish
emission from $\sim$million K gas.

\par
It is also possible that the differences between the two
regions are the result of rapid cooling.
Non-uniformities in the density field, or in the heating
mechanism may accentuate differences in cooling across the
X-ray plume, and lead to a more clumpy gas distribution
\citep[as in, e.g.,][]{Nulsen1986}.
Indeed, the granularity of the sub-keV gas around NGC~4696
(see fig.~1 in C11a) is reminiscent of a collection of clouds.

\par
On the other hand, \citet{Panagoulia+2013-cen-iron} reported
recently a significant drop in iron abundance within
5--10\unit{kpc} of the nucleus in NGC~4696, which is of the
order of a few relative to the more remote (say, 20\unit{kpc}) ICM.
The possibility that the north-west region observed by C11a lies
within the depleted region would provide a natural explanation to
the absence of any detectable coronal iron emission.
It would also require that the X-ray plume is elongated along the
line of sight, so that the south-east region lies farther out,
and is not drastically affected by the abundance drop.

%%%%%%%%%%%%%%%%%%%%%%%%%%%%%%%%%%%%%%%%%%%%%%%%%%%%%%%%%%%%%%%%%%%%%%%%%%%%%%%%%
%				FIGURE						%
%%%%%%%%%%%%%%%%%%%%%%%%%%%%%%%%%%%%%%%%%%%%%%%%%%%%%%%%%%%%%%%%%%%%%%%%%%%%%%%%%
%%%%%%%%%%%%%%%%%%%%%%%%%%%%%%%%%%%%%%%%%%%%%%%%%%%%%%%%%%%%%%%%%%%%%%%%%%%%%%%%%

\section{Summary}\label{sec:summary}

\par
We have employed self-consistent non-equilibrium cooling calculations
implemented in \cloudy{} to study the thermal properties of the coronal 
line gas near the core of NGC~4696.
The non-detection of the \emlineforb{Fe}{xiv} line suggests that the
coronal cloud is not a cooling condensation of the surrounding ICM.
Fine-tuned values of intrinsic extinction would be required to explain
the detection of the optically thin \emlineforb{Fe}{x}, but not of
\emlineforb{Fe}{xiv}.

\par
The failure to detect \emlineforb{Fe}{xiv} suggests a temperature ceiling
for the coronal cloud, which we estimate at $\sim$2.1~million K.
This result is not sensitive to the presence of a temperature floor.
The coronal gas is likely embedded in the X-ray plume, but even if it is not,
it cools isochorically, and its mass is $\sim$million {$M_\odot$}.

\par
Finally, we have briefly explored a few scenarios for the origin of the
coronal cloud, and we have pointed out that the cloud is consistent with
being either the conductive or the mixing interface between the hot ICM
and the warm optical gas.
Detailed calculations for observable signatures of these models are
beyond the scope of the paper.
None the less, we have presented a number of bright
lines that trace gas down to 10$^4$\unit{K}.

\section{Acknowledgements}
We would like to thank Orly Gnat \& Matthias Steffen for valuable
assistance with the validation of the time dependent scheme in \cloudy.
GJF acknowledges support by NSF (0908877; 1108928; and 1109061), 
NASA (10-ATP10-0053, 10-ADAP10-0073, and NNX12AH73G), JPL (RSA No 1430426),
and STScI (HST-AR-12125.01, GO-12560, and HST-GO-12309).
ACF thanks European Research Council for the Advanced Grant FEEDBACK.
PvH acknowledges support from the Belgian Science Policy Office
through the ESA PRODEX program.

\bibliographystyle{mn2e}
\bibliography{bibliography2,coronal}

\appendix
\section{\cloudy{} input}

\par
As a means to reproduce our results, we provide a minimal input script
for isobaric cooling in collisional ionization equilibrium.
{\footnotesize
\begin{verbatim}
coronal 8e7 K init time
set dynamics population equilibrium
iterate to convergence
stop time when temperature falls below 1e4 K
c
atom chianti "CloudyChiantiAll.ini"
hden 1 linear
constant gas pressure reset
c
set dr 0
set zone 1
\end{verbatim}
}

\bsp

\label{lastpage}
\clearpage
\end{document}

%% file: macros.tex
\font\manual=manfnt at 7pt \def\dbend{\hbox{\raise0.9ex\hbox{\manual\char127\hspace{0.6em}}}}

\newcounter{INTERNALionstage}

\def\gtsim{\mathrel{\hbox{\rlap{\hbox{\lower4pt\hbox{$\sim$}}}\hbox{$>$}}}}
\def\lesssim{\mathrel{\hbox{\rlap{\hbox{\lower4pt\hbox{$\sim$}}}\hbox{$<$}}}}

\def\ga{{\rm\thinspace gauss}}

\def\la{\mbox{{\rm L}$\alpha$}}

\def\ha{\mbox{{\rm H}$\alpha$}}
\def\hb{\mbox{{\rm H}$\beta$}}
\def\hi{\mbox{{\rm H~{\sc i}}}}

\def\htwo{\mbox{{\rm H}$_2$}}

\DeclareMathAlphabet{\vib}{OML}{cmm}{m}{it}
\newcommand*{\satellite}[1]{\textit{#1}}
\newcommand*{\xmm}{\satellite{XMM-Newton}}
\newcommand*{\chandra}{\satellite{Chandra}}

%% file: LocalMacros.tex
\newcommand{\cloudy}{{\sc cloudy}}

\newcommand{\unit}[1]{\ensuremath{\mathrm{\thinspace #1}}}
\newcommand{\Mdotunit}{\ensuremath{M_\odot \, \mathrm{yr^{-1}}}}

\newcommand{\emlineforb}[2]{\ensuremath{[}#1 {\sc #2}\ensuremath{]}}
\newcommand{\emline}[2]{#1 {\sc #2}}